# The Link between Magnetic-field Orientations and Star Formation Rates

Hua-bai Li, Hangjin Jiang , Xiaodan Fan, Qilao Gu & Yapeng Zhang
The Chinese University of Hong Kong

**Understanding star formation rates (SFR) is a central goal of modern star-formation models, which mainly involve gravity, turbulence and, in some cases, magnetic fields (B-fields)[1,2]. However, a connection between B-fields and SFR has never been observed. Here, a comparison between the surveys of SFR[3,4] and a study of cloud–field alignment[5]—which revealed a bimodal (parallel or perpendicular) alignment—shows consistently lower SFR per solar mass for clouds almost perpendicular to the B-fields. This is evidence of B-fields being a primary regulator of SFR. The perpendicular alignment possesses a significantly higher magnetic flux than the parallel alignment and thus a stronger support of the gas against self-gravity. This results in overall lower masses of the fragmented components, which are in agreement with the lower SFR.**

It is not difficult to imagine that the SFR of a molecular cloud should be related to the mass of gas it contains, due to self-gravity. On the whole, this trend has indeed been observed[3,4]. On the other hand, star formation efficiencies of molecular clouds are usually just a few percent[6] while cloud ages are at least comparable to their free-fall time (~Myr)[7]. This requires other forces to slow down gravitational contraction[8]. In addition, clouds with a similar mass and age can have different SFR; for example, it is well known that the Ophiuchus cloud has a significantly higher SFR than its neighbour, the Pipe Nebula[3,4]. In Figure 1, we can see a significant difference between the two dark clouds: Ophiuchus is accompanied by the colourful nebulae, which is a signature of active stellar feedback, while Pipe looks very quiescent in comparison. No one knows the reason for these differences in SFR. Another good example is the pair of Rosette and G216-2.5 molecular clouds[9], where turbulence had been suspected as the cause of their very different SFRs. But later their turbulent velocity spectra were found almost identical[9].

Recently, a good agreement has been discovered between the empirical column density threshold for cloud contraction and the magnetic critical column density of the Galactic field (~10 µG)[5,10]. For densities lower than this threshold, the field strength is independent of densities[11]; i.e., gas must accumulate along field lines. Also, a bimodal cloud-field alignment, which is another signature of field-regulated (sub-Alfvenic turbulence) cloud formation, was recently observed in the Gould Belt[5], where, interestingly, Pipe and Ophiuchus (Figure 1) are aligned differently from their local fields. More intriguingly, it is the one aligned with the B-field that holds the higher SFR. Together with another observed piece of evidence that Galactic B-field direction anchors deeply into cloud cores[12] and thus plays a role in cloud fragmentation[13], we were prompted to survey whether the cloud–field alignment has a connection with SFR.

Luckily, both the cloud–field alignment[5] and SFR[3,4] of the Gould Belt clouds have been very well studied (Table 1). Both Heiderman et al.[3] and Lada et al.[4] assume that SFR is directly related to the population of young stellar objects (YSOs) in a cloud. The YSOs in the Gould Belt have similar ages of 2 ± 1 Myr (millions of years) and the median for the initial stellar mass is around 0.5 $M_\odot$ (solar mass), so the SFR of each cloud can be estimated by the number of embedded YSOs multiplied by 0.5 ($M_\odot$) and divided by 2 (Myr). As mentioned at the outset, we are looking for factors in SFR other than mass, so here we study SFR per unit mass (SFR/mass). Heiderman et al. estimate cloud mass from regions above $A_v = 2$ mag, and Lada et al., besides a threshold similar to Heiderman et al., also used $A_v = 7$ mag. Note that while the Herschel space telescope revealed ubiquitous sub-cloud filamentary density structures[14], where most of the protostars are forming, YSOs, on the other hand, do not correlate with these filaments in position[15,16] (see Supplementary Figure 1 for example). So the resolution of SFR[3,4], and thus our study of their correlation with field orientations, remains at cloud scales.

Dust grains in molecular clouds cause the 'extinction' of background starlight (see Figure 1 for examples), which is widely used to study density structures and B-fields at cloud scales. A cloud orientation can be specified by the direction in which the autocorrelation length of the extinction map reaches the maximum[5]. The residual background starlight after extinction is polarized along the B-fields. Li et al.[5] defined a cloud B-field direction by the Stokes mean of all the polarization detections in the ~10-parsec region surrounding the cloud. Based on Monte Carlo simulations and Bayesian analysis[5], they concluded that 95% of the 3-D cloud-field angles must be within 20º from either 0º or 90º.

Figure 2 correlates the SFR/mass with the cloud–field angles and shows that large-angle clouds have consistently lower SFR/mass, no matter which $A_v$ threshold is used. Here we perform hypothesis tests to see how significant the trend is. Due to the small sample size, a non-parametric permutation test is more appropriate than a parametric test.

Let $\mu_L$ and $\mu_S$ be the population mean SFR/mass of the large-angle group, L, and the small-angle group, S, respectively. To test whether $\mu_L$ is significantly smaller than $\mu_S$, we adopt the null hypothesis $H_0$ : L and S stem from the same population distribution, i.e., $\mu_L = \mu_S$, and the alternative hypothesis $H_1$ : $\mu_L < \mu_S$. As the sample mean is an unbiased estimator of the population mean, we use $T_{obs}$ = <L> − <S> as the test statistic under observed samples; where <...> means the sample mean. The logic of the permutation test goes as follows. Under the null hypothesis, Z = (L, S), the combination of L and S, should still be an independent and identical sample from the same population distribution. Any regrouping of Z into two groups, $Z_1$ and $Z_2$, of the same sizes as L and S will also produce two independent samples from the same distribution. The test statistics, T = < $Z_1$ > − < $Z_2$ >, from such kind of regroupings should have a mean equal to 0. By enumerating all possible regroupings of Z while maintaining the same sample sizes, we will obtain $^{m+n}C_m$ ("m+n choose m") values of T's, where m and n are the sample sizes of L and S. The percentage of the T's that are lower than or equal to $T_{obs}$ is an estimate of the p-value. $H_0$ should be rejected if the p-value < 0.05.

The data from Lada et al. has m = n = 4 and the $T_{obs}$ = 3.66 % Myr$^{-1}$ (Table I) is lower than T from any regrouping. This is not a surprise given that L is composed of the four lowest SFR/mass in Z. So the p-value is $1/^8C_4 = 0.014$. The Heiderman et al. data comes with the estimates of uncertainties (Table I) and we will take the uncertainty into consideration based on a permutation test on the bootstrap estimates of the test statistics. For the given mean and standard deviation of the SFR/mass observed from a cloud, we treat them as the mean and standard deviation of a log-normal distribution of SFR/mass. For a grouping of Z, to get a bootstrap estimate of the test statistic, one value for each cloud is randomly sampled from the corresponding log-normal distribution, and from these sampled values, a test statistic, $T_i$, can be derived. After 1000 rounds of the above random sampling, 1000 $T_i$'s are obtained, and their mean is used as the estimate of the test statistic of that particular grouping. The same process is applied to all the $^{11}C_5$ groupings, and the test statistic derived from the grouping of L and S gives the estimate of $T_{obs}$. Accordingly, the p-value is 0.0043.

Besides the permutation tests, Spearman rank correlation tests are also performed and summarized in Methods, which show that the negative correlation between SFR and cloud-field angles (Figure 2) is also significant (p-value < 0.006). Also in Methods is the analysis of Planck polarimetry data as a double-check of our discovery. Three reasons motivate Li et al.[5] to use optical data. First, both observations[12,13] and simulations[17,18] have shown that field directions from low-density cloud vicinities, where optical data tends to trace, are closely aligned with fields inside the clouds down to cloud cores. Second, for cloud vicinities, Planck data suffers more background contamination, because the distances of background stars can be selected for optical polarimetry analysis but Planck is sensitive to the entire lines of sight. Third, cloud B-fields have a higher chance of being affected by the embedded stellar feedback than the fields in the vicinities, so the latter should better preserve mean field orientations prior to star formation; a good illustration is Perseus, as discussed in Methods. However, given the first point, stellar feedback should not completely change field morphologies, and we should still see some correlation between Planck data and SFR. Indeed, hypothesis tests (see Methods) suggest that Planck and optical data correlate with SFR similarly, which, given the similar trends seen in Figure 2, should not be difficult to conceive. While the Planck team[30] show that the alignment between column density contours and B-fields tends to move away from parallelism as column density increases, we stress that it should not be interpreted as B-fields tending to be perpendicular to high-density structures ($A_V > 2$ mag), as we see comparable populations for large and small cloud-field angles (Figure 2; reference 5). Supplementary Figure 5 is dedicated to bridging the analyses from reference 5 and 30.

So SFR/mass is observed to be significantly lower for large-angle clouds. Though the fact that cloud fields are ordered has only been discovered very recently[10,12,13], the way in which stars can form under this condition has been considered for more than half a century: the mass-to-magnetic-flux ratio, $M/\Phi$, has to be above a critical value for gas to collapse against magnetic pressure[19,20]. Here we try to understand Figure 2 along the same line of thought.

First, the $M/\Phi$ of an elongated cloud depends on cloud-field orientation. To visualize this, consider a cylindrical cloud with length $\ell >$ width $d$ (Figure 3). If the long axis is aligned parallel or perpendicular to B-fields, the magnetic flux is, respectively, $B\pi d^2/4$ and $B\ell d$ (Figure 3). The ratio between the two is in the order of $d/\ell < 1$, i.e. when aligned perpendicular to B-fields, a cloud as a whole experiences a higher magnetic flux and thus more support from B-field against self-gravity. In other words, the orientation parallel to the field possesses a higher "magnetic criticality" (the ratio of $M/\Phi$ to the critical value), which agrees with the overall higher SFR/mass.

We can further test the idea as follows. For the $M/\Phi$ of a subregion, consider a uniform linear density, $\lambda$, along the cloud long axis. For the parallel alignment, $M/\Phi$ of a subregion is proportional to the scale, s, along the axis: $4\lambda s/B\pi d^2 \simeq \lambda s/Bd^2 = (s/d)\lambda/Bd$. On the other hand, for the perpendicular case, $M/\Phi = \lambda/Bd$ is independent of s. Only the $M/\Phi$ of parallel alignment can exceed $\lambda/Bd$ and grow a massive fragment that is impossible in the perpendicular case. Kaufmann et al.[21] studied the mass-size relation of the fragmentation in clouds Taurus, Perseus, Pipe, and Ophiuchus, which can serve as a test of the argument above. Their results are summarised in Figure 3 (right panel): the upper limit of the fragmented mass is indeed systematically higher for the parallel case.

For subregions with $M/\Phi$ below the critical value, fragmentation can still be achieved by either reducing the flux through ambipolar diffusion or increasing the mass through gas accumulation along the fields[19,20]. Whether the former presents in molecular clouds is still under intense debate (e.g. reference 22-24; also, field diffusion due to other mechanism - magnetic reconnection - is proposed recently[25]). In the latter scenario, a smaller cloud-field angle is easier for both gravity and turbulence to accumulate gas along the field to enhance $\lambda$ (and thus the subregion $M/\Phi$).

Of course, in reality, clouds are not uniform cylinders and our model is still far from making a quantitative prediction, but the assumption helps us appreciate the fact that B-fields perpendicular to a cloud can more effectively hinder massive fragmentation and thus SFR. Magnetic fields are generally treated as isotropic pressure, like gas pressure, in existing SFR models[1,2]. However, the anisotropy of magnetic pressure is why the cloud-field alignment matters in the discussion above. As the discovery of ordered cloud B-fields has just started to constrain theories and numerical simulations of star formation (e.g. reference 10,17,18,26,27), being able to explain the bimodal cloud-field alignment[5] and SFR (Figure 2) should also be included among the criteria of a successful cloud/star-formation theory.

## Methods

*Planck polarimetry data*

For each Gould-Belt cloud, Planck Collaboration[30] defined the lowest, the intermediate, and the highest $N_H$ bin. We simply adopt their thresholds for the intermediate density, which is $N_H = 10^{21.5 \pm 0.17}$ cm$^{-2}$ (see examples in Supplementary Figures 2), to define the regions for which we calculate cloud mean fields. We stay away from low-density regions because they are more vulnerable to background contamination (see discussion in main text). Also, these thresholds of intermediate density are quite close to the ones used to define cloud masses in the SFR study[3] (Supplementary Figures 1). The 14 regions for mean field calculation are shown in Supplementary Figures 1- 3 and the results are listed in Supplementary Table 1. In a similar way to reference 5, the mean field direction can be estimated by utilizing the Stokes mean of all the polarized detections within a selected region.

*Comparison between Planck and optical polarimetry data*

In reference 5, for each cloud, optical polarimetry data are collected from a region > 10 pc in scale in order to get enough detections for the mean field direction. If two clouds are too close, e.g. Musca/Chamaeleon and Orion A/B, reference 5 cannot resolve them (see footnote "d" and "g" of Table 1). There is no such limitation given to the 5-arcmin resolution of Planck, so there are two more clouds in Supplementary Table 1.

The most prominent difference between the two data sets occurs with Perseus cloud: it switches from a small-angle cloud to a large-angle cloud (Figure 2). Goodman et al.[31] found that the B-field direction inferred from optical polarization in the Perseus region is bimodal (Supplementary Figure 4), with less polarized vectors (P < 1.2%) aligned along the cloud's long axis and stronger polarized vectors (P > 1.2%) lying roughly perpendicular to the long axis. Reference 5 took an equal-weight mean of all the vectors, so the group with more vectors, which is the one aligned with the cloud, wins. Planck data, on the other hand, favours the regions with higher polarized flux along a line of sight, which can be more comparable to the stronger polarized optical vectors. Since the polarized ratio may increase with the distances of stars[32], Goodman et al.[31] proposed that two clouds at different distances possessing differing field orientations could be superimposed along the line of sight.

The background of Supplementary Figure 4 is IRAS-derived dust column density[33], which traces primarily warm dust surrounding young stars and H$_{II}$ regions. Since the polarization efficiency of dust associated with cold molecular gas can be lower than the efficiency of dust embedded in warm atomic gas[34], Bally et al.[33] proposed that the stronger polarization in Perseus is predominantly associated with the ionization front of the H$_{II}$ region G159.6-18.5, which is well traced by the IRAS-derived dust map, and the less polarized vectors mostly trace magnetic fields associated with the Perseus cloud itself.

It is also common to see that thermal dust emission polarimetry data traces B-fields associated with ionization fronts[13,32]. Further away from the H$_{II}$ region, we can see in Supplementary Figure 4 that the fields from NGC 1333 4A/B traced by sub-mm polarimetry are also parallel with the long axis of Perseus.

In any case, the situation of Perseus is unique, so the permutation tests (see main text) are repeated using Planck data either with or without Perseus. The results are as follows:

| Planck p-value | SFR Lada et al. | SFR Heiderman et al. |
|---|---|---|
| with Perseus | 0.056 | 0.0065 |
| without Perseus | 0.014 | 0.004 |

Therefore, the bimodal SFR is still significant with Planck data.

*Spearman rank correlation test*

While reference 5 has concluded a 3-D bimodal cloud-field alignment, and we adopted this point of view to show that SFR/mass can also be divided into two groups, it is difficult to distinguish "two clusters" from "one decreasing trend" for the plots in Figure 2. The mass-to-flux model we proposed to explain Figure 2 is consistent with both possibilities. So here we study how significant the decreasing trend is.

We use Spearman rank correlation (SRC) test, which is based on the Pearson Correlation Coefficient between the ranks of two marginal observations. Since the SRC test is based purely on ranks instead of the original quantities, it is insensitive to outliers. For our cases, SRC(rank$_{SFR}$, rank$_{cf}$) are between the ranks of SFR, rank$_{SFR}$, and the ranks of cloud-field angles, rank$_{cf}$; they are as follows:

| SRC (p-value) | SFR Lada et al. | SFR Heiderman et al. |
|---|---|---|
| Optical | -0.83 (0.006) | -0.75 ($\approx 0$) |
| Planck | -0.61 (0.043) | -0.81 ($\approx 0$) |

SRC ranges within ± 1; the more positive/negative a SRC is, the more positively/negatively correlated are the ranks. SRC ≈ 0 means no correlation. Therefore, a negative correlation between SFR and cloud-field angles, i.e. a decreasing trend in Figure 2, is clearly based on the observed SRCs (SRC$_{obs}$) in the above table. In a similar way to the permutation tests for the bimodality, we perform a permutation test for the decreasing trend. For the SFR from Lada et al.[4], we randomly permute rank$_{SFR}$ to generate rank$_{SFR\_rand\_i}$, for i = 1, 2, ..., k, where k is the total number of permutations used for p-value estimation. We adopt the null hypothesis H$_0$ : SRC(rank$_{SFR}$, rank$_{cf}$) = 0, and the alternative hypothesis H$_1$ : SRC(rank$_{SFR}$, rank$_{cf}$) < 0. The p-value is thus estimated by the relative frequency to obtain SRC(rank$_{SFR\_rand\_i}$, rank$_{cf}$) < SRC$_{obs}$. The p-values in the table above are estimated with k = 1000.

For the SFR and their uncertainties from Heiderman et al.[3], again, we treat them as the means and standard deviations of log-normal distributions. To obtain a bootstrap sample, we sample one value for each cloud from their corresponding log-normal distribution. In a similar way to the Lada et al. data, a p-value can be estimated by permutation for this particular bootstrap sample. By repeating this sampling N times, we obtained N resampling p-values (p$_1$, p$_2$, ..., p$_N$), which resulted from N independent and identically distributed samples. Let $S_{obs} = -2 \sum_{t=1}^{N} \log(p_t)$, S$_{obs}$ is known to follow a chi-square distribution with degree of freedom 2N if H$_0$ is true. Thus the p-value is estimated by the right-tail probability P($\chi^2_{2N}$ > S$_{obs}$). The decreasing trends are highly significant based on the p-values reported in the above table, where N=1000.

*Data availability*
The observational data that support the plots within this paper and other findings of this study can be found from references 3-5 and 30. The Planck 353 GHz data can be downloaded from http://pla.esac.esa.int/pla/#maps.

Table 1 Cloud long axis directions, B-field directions and the ratio of SFR to cloud mass.

| Cloud Name | Long axes[a] degrees | B fields[a] degrees | SFR/cloud mass[b] % Myr$^{-1}$ | SFR/cloud mass[c] % Myr$^{-1}$ |
|---|---|---|---|---|
| 1. IC 5146 | -38 | 64 ±16 | | 0.380±0.18 |
| 2. Pipe Nebula | -45 | 49 ±13 | 2.81 | |
| 3. Orion | 90 | 1± 27 | 4.17[d] | |
| 4. Chamaeleon | 19 | -71± 11 | | 1.03±0.48 (0.900±0.56)[g] |
| 5. Taurus | 75 | 0 ± 18 | 4.76 | 0.140 |
| 6. Lupus I | -1 | -82 ± 13 | 4.00 | 0.630±0.52 |
| mean of 1-6 | | | 3.94±0.82 | 0.620±0.37 |
| 7. Lupus II-VI | -73 | 81 ± 11 | 6.97[e] | 1.85±0.74[f] |
| 8. Corona Aus. | -26 | -27 ± 32 | 9.69 | 3.66±2.4 |
| 9. Cepheus | 65 | 69 ± 36 | | 1.13±0.62 |
| 10 Ophiuchus | -45 | -65 ± 25 | 6.10 | 2.32±1.8 |
| 11. Aquila | -75 | -45 ± 10 | | 1.48±0.8 |
| 12. Perseus | 32 | 59 ± 35 | 7.98 | 1.46±1.1 |
| mean of 7-12 | | | 7.60±1.5 | 1.98±0.92 |

a. Cloud long axes and B-field directions adopted from reference 5 (increasing counterclockwisely from Galactic north). The first 6 clouds have directional differences which are larger than 70 degrees, while the other 6 have a difference which is less than 30 degrees. Errors related to field direction errors are defined by the interquartile ranges of the polarization detections. The long-axis direction errors are less than 15 degrees[5].
b. SFR and cloud mass adopted from reference 4, from which errors are not available.
c. SFR and cloud mass adopted from reference 3; errors are propagated from those of SFR and cloud mass. The SFR/mass from reference 3 is systematically lower than those from reference 4, since the latter used a larger column density threshold to estimate the cloud masses.
d. Orion A & B.
e. Lupus 3 & 4.
f. Lupus 3-6.
g. For Musca, which is included in Chamaeleon in reference 5.

Figure 1 - **Filamentary clouds and B-fields in the Pipe-Ophiuchus region**. The dark lanes are clouds traced by the extinction of the background starlight (photo credit: Stéphane Guisard). The line segments indicate B-field directions inferred from the polarization of the background starlight. Pipe Nebula, overlapped with the red lines[28], tends to be perpendicular to the B-fields. On the other hand, Ophiuchus cloud, covered by the yellow lines[29], is largely aligned with the fields. They are not special cases from the Gould Belt, where most molecular clouds are elongated and tend to be aligned either parallel or perpendicular to local B-fields[5].

Figure 2 - **SFR per unit mass versus cloud-field alignment for the Gould Belt clouds**. The SFR and cloud mass are adopted from the surveys performed by Heiderman et al.[3] and Lada et al.[4] (Table I). For each survey, the SFR/mass is normalized to the mean. The cloud long axes and B-field directions based on optical data in the upper plot are adopted from Li et al.[5] (Table I). In the lower plot, we replace the B-field directions from Li et al.[5] with Planck 353 GHz polarization data[30] (Methods; Supplementary Table 1). Perseus cloud, shown as hollow symbols, has significantly different mean field directions based on optical and Planck data (see the main text and Methods for discussion).

The "alignment parameters" used in reference 30 are shown in Supplementary Figure 5 for the clouds noted as Ⓐ (for aligned) and Ⓟ (for perpendicular) to illustrate the connection between the two analyses.

Figure 3 - ***Left*: An illustration of how the magnetic flux of an elongated cloud can vary with the cloud-field alignment.** A cross-section perpendicular to the B-field is highlighted for the cylindrical cloud when it is parallel (red) or perpendicular (green) to the local field. The area of each cross-section is shown below the cloud. ***Right*: The distribution of mass-size relation, m(r), of the fragmentation from four Gould-Belt clouds**[21]. Comparable to the left panel, the areas within the red dashed line or red shaded region are from small-angle clouds, based on the optical polarimetry data. The large-angle clouds, within the green dashed line or green shaded region, have an overall lower fragmented mass. The dotted dark lines are for references, which follow either $m(r) \propto r^2$ or $m(r) \propto r$ for an isothermal equilibrium sphere at 10K.

**Acknowledgements**
The research was supported by Hong Kong Research Grant Council, projects T12/402/13N, ECS24300314 and GRF14600915; also by CUHK Direct Grant for Research, project 4053126 'Analyzing Simulation Data of Star Formation'. H.L. appreciates the conference "Star Formation in Different Environments 2016", where the discussion, especially with Christopher Matzner and Juan Diego Soler, inspired the direction we present this work. Q.G. thanks Yi Wang for the discussion on Planck data analysis.

**Author contributions**
H.L. designed and executed the experiment. H.J. and X.F. were in charge of the statistical tests. Q.G. and Y.Z. were responsible for the Planck data analysis.

**Competing financial interests**
The authors declare no competing financial interests.

**Corresponding author**
Hua-bai Li

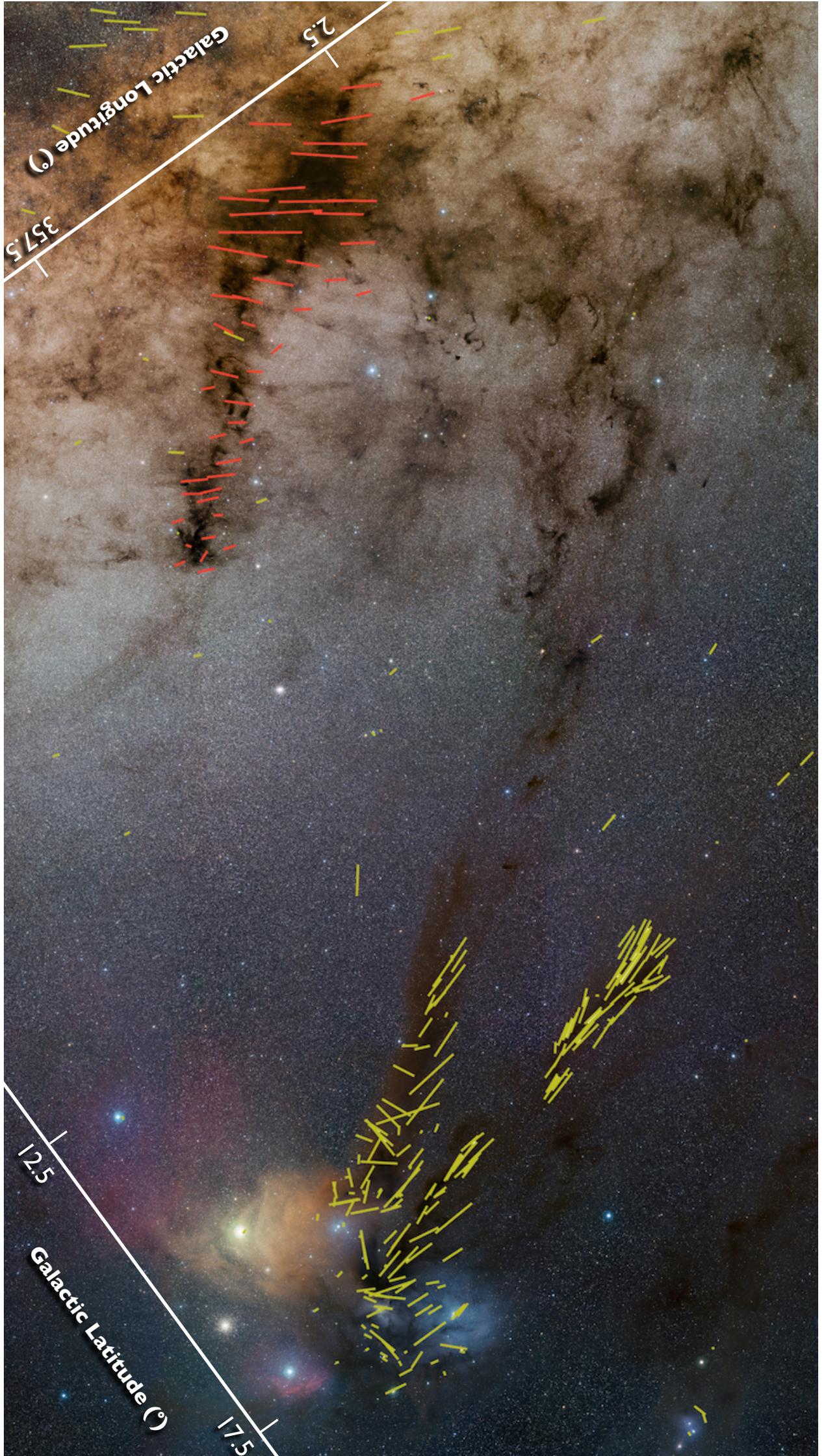

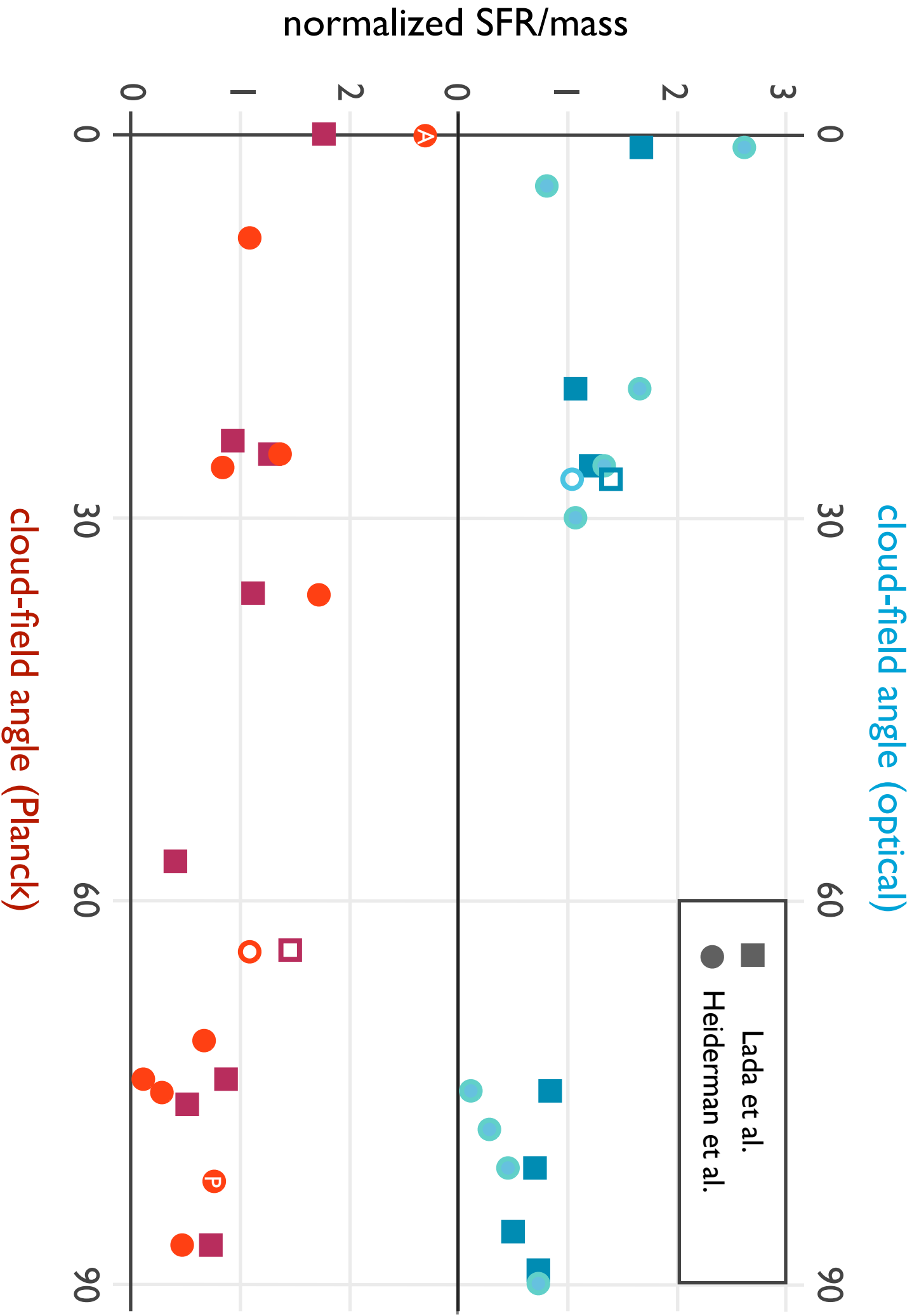

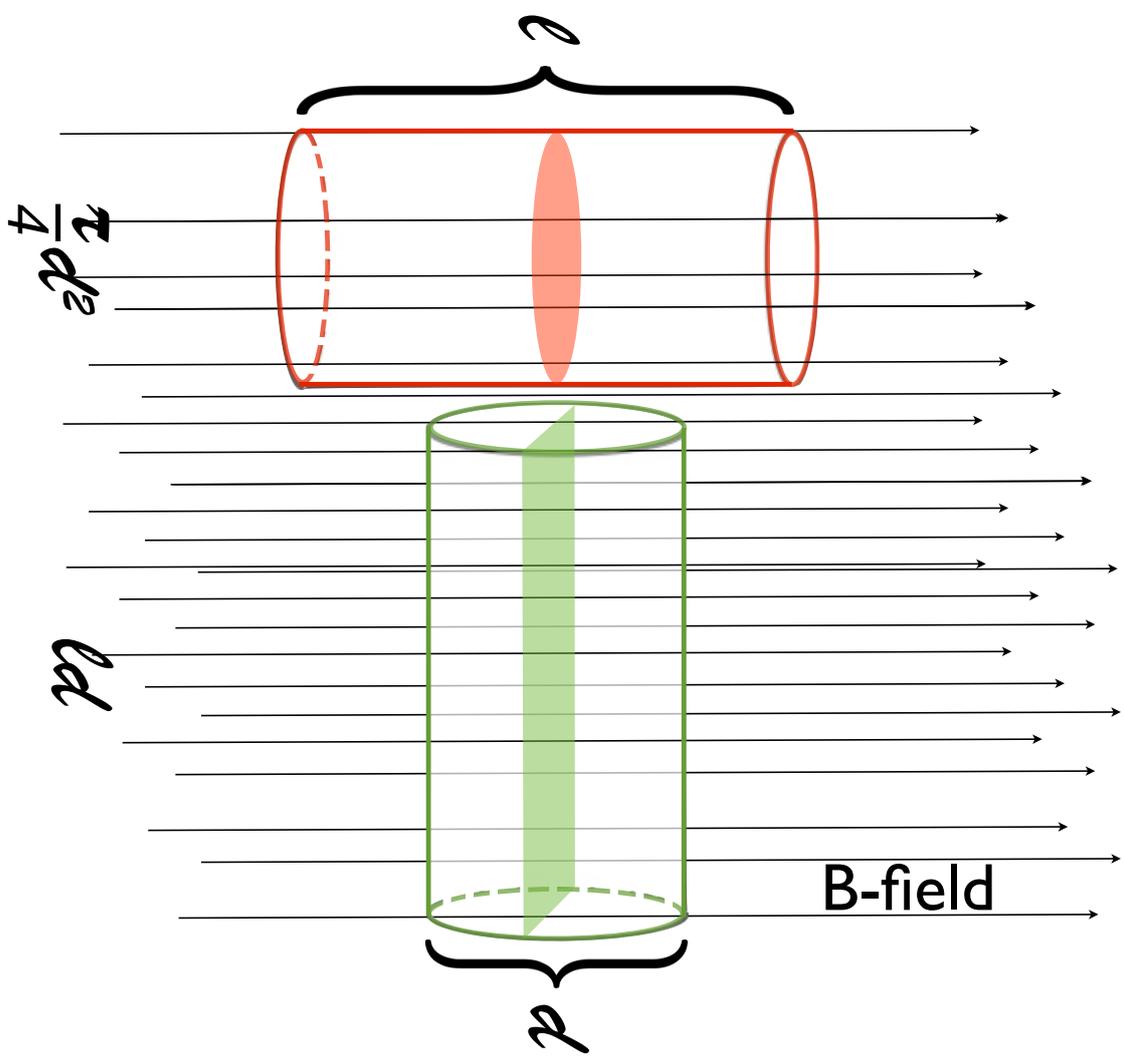
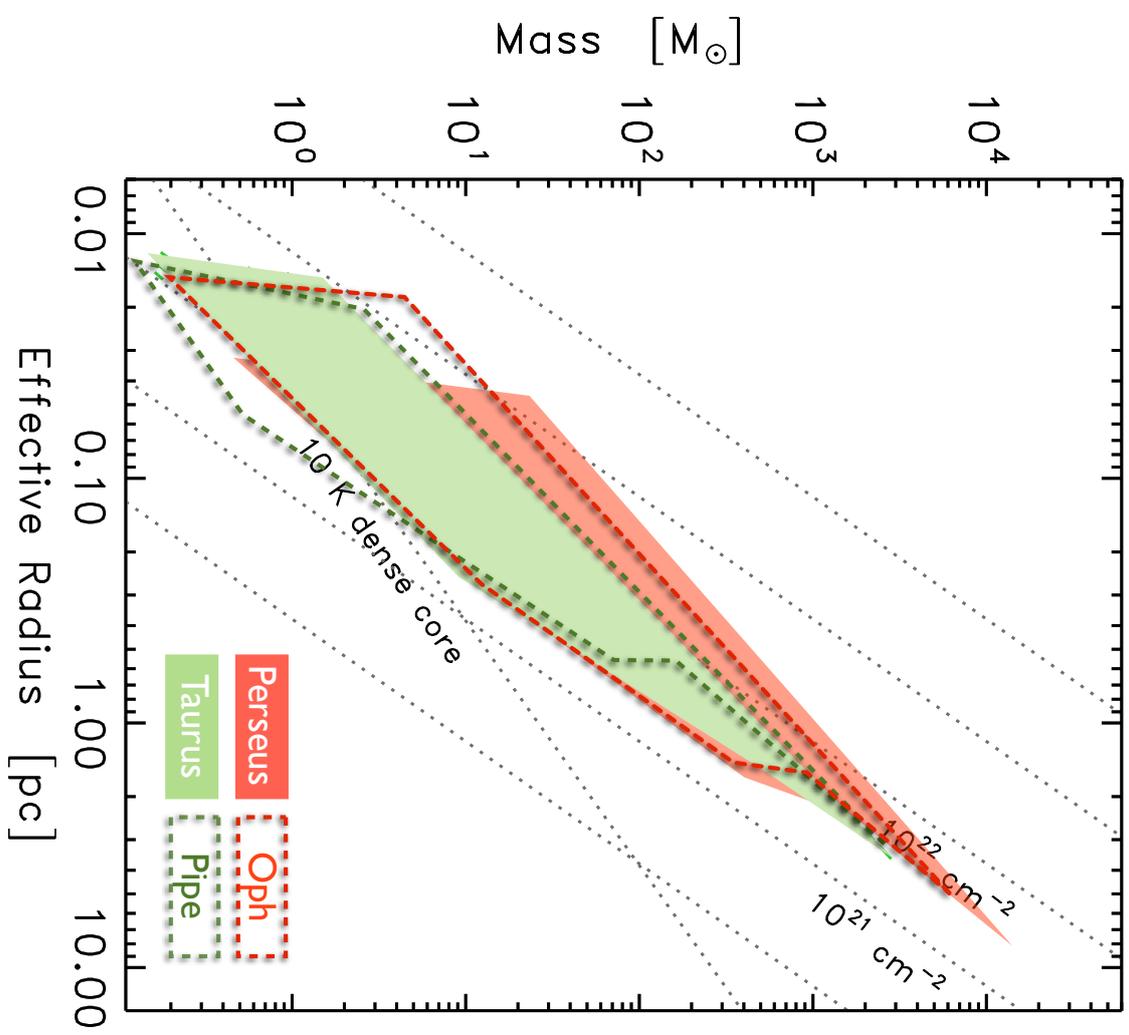

# Supplementary Information

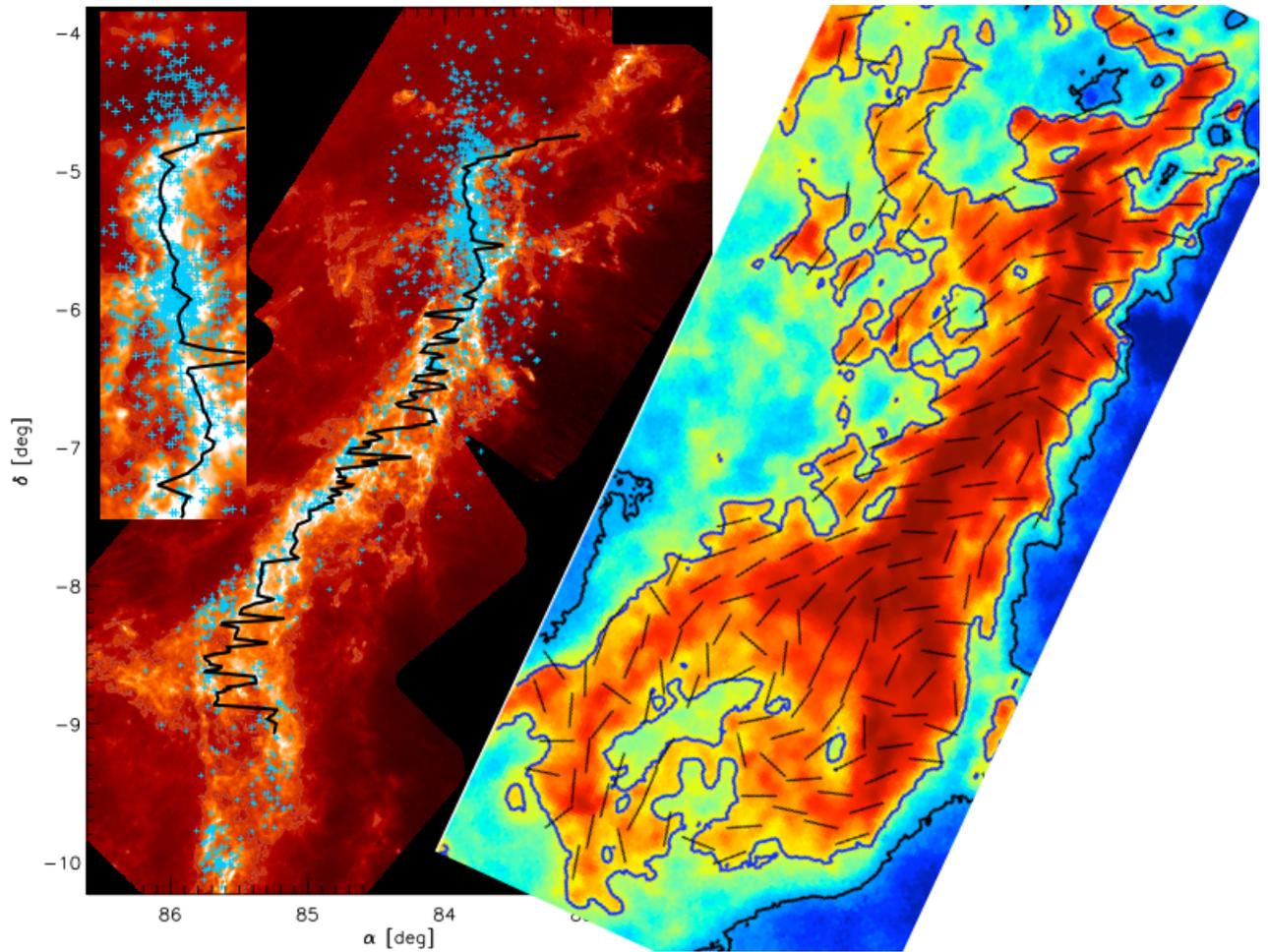

**Supplementary Figure 1 - YSOs (blue dots in the left panel[16]) and B-field directions (vectors in the right panel[30]) of Orion A.** The map of YSOs is adopted from Stutz & Gould[16], who concluded that YSOs, unlike protostars, are essentially uncorrelated with fine filaments within the cloud. The overall distribution of YSOs roughly correlates with the bulk volume above $A_v$ = 2 mag, which is used by reference 3 to estimate cloud masses for SFR. The vectors in the right panel are also above a similar threshold (the blue contour; $N_H = 10^{21.61}$ cm$^{-2}$).

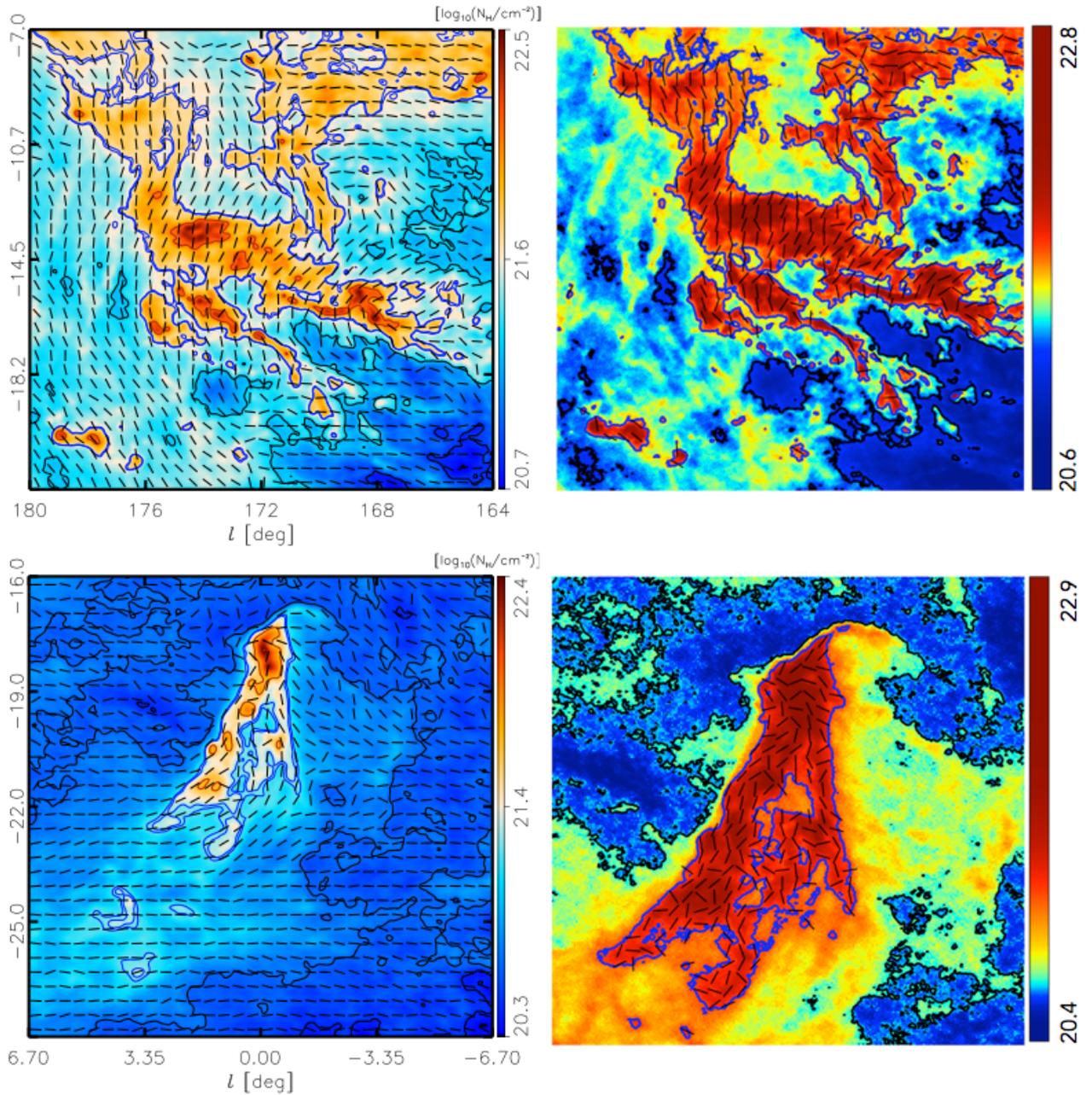

**Supplementary Figure 2 – An illustration of how the polarization data from reference 30 is adopted in this study.** *Left*: Adopted from Planck Collaboration[30] are the column density maps, $\log_{10}(N_H/cm^2)$, overlapped with magnetic field directions (vectors) inferred from the 353 GHz polarization data. Taurus and CrA are shown as examples. The thresholds of intermediate density[30] are indicated by the blue contours. *Right*: We collected the Planck 353 GHz polarization data above the intermediate-density thresholds and calculated the mean B-field directions. Another example, Orion A, is shown in Supplementary Figure 1 (right panel). The maps of the rest 11 clouds are given in Supplementary Figure 3.

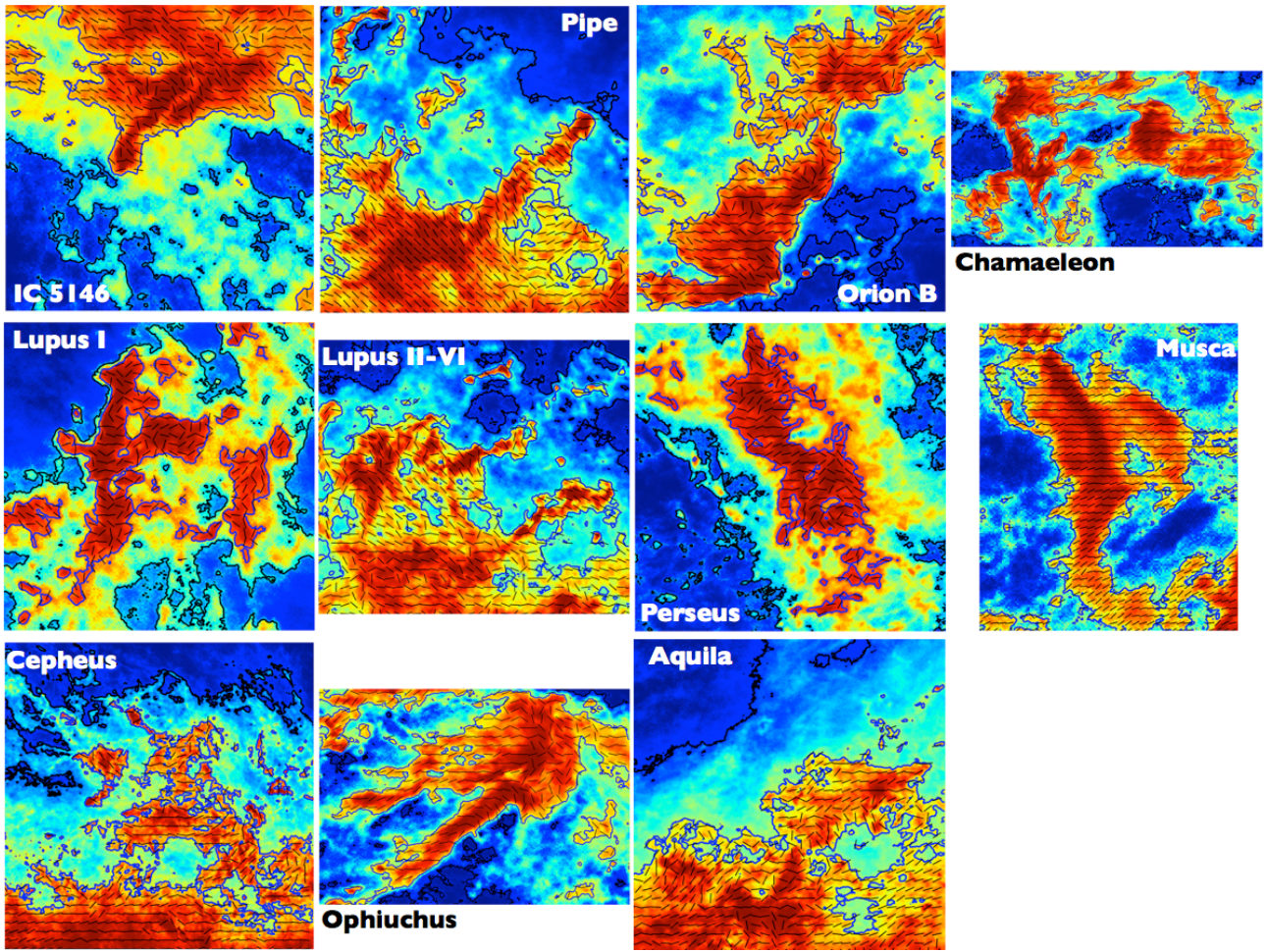

**Supplementary Figure 3 - All the regions from which we derive cloud B-fields based on the Planck data**[30], except for the three clouds already shown in Supplementary Figures 1 and 2. The regions are above the Planck-defined intermediate-density thresholds[30].

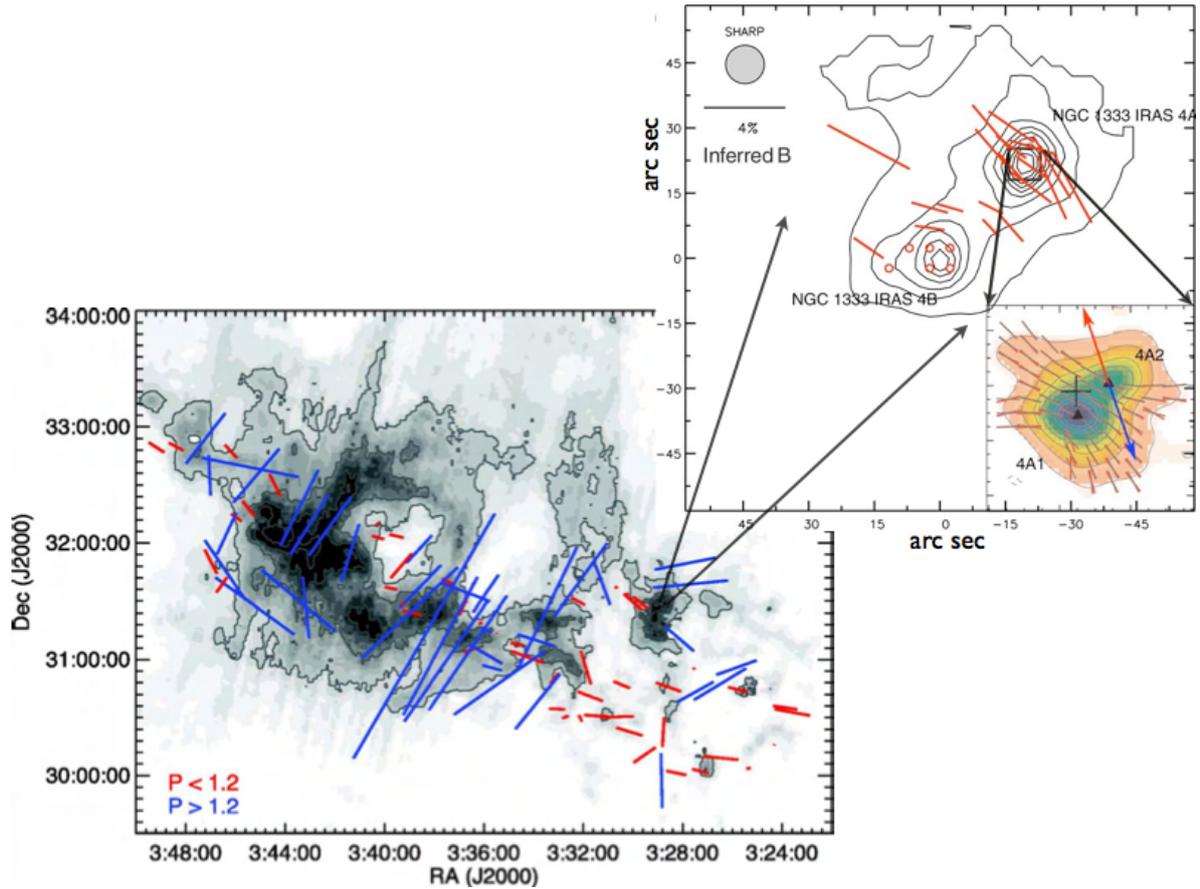

**Supplementary Figure 4 - Optical polarization vectors on top of an IRAS-derived dust map of Perseus.** The figure is adopted from Bally et al.[33] to show the bimodal field direction and the bimodal degree of polarization (indicated by the vector length). CSO[35] and SMA[36] zoomed-in the NGC 1333 4A/B cores (upper-right panel) and revealed cloud B-filed embedded in high densities.

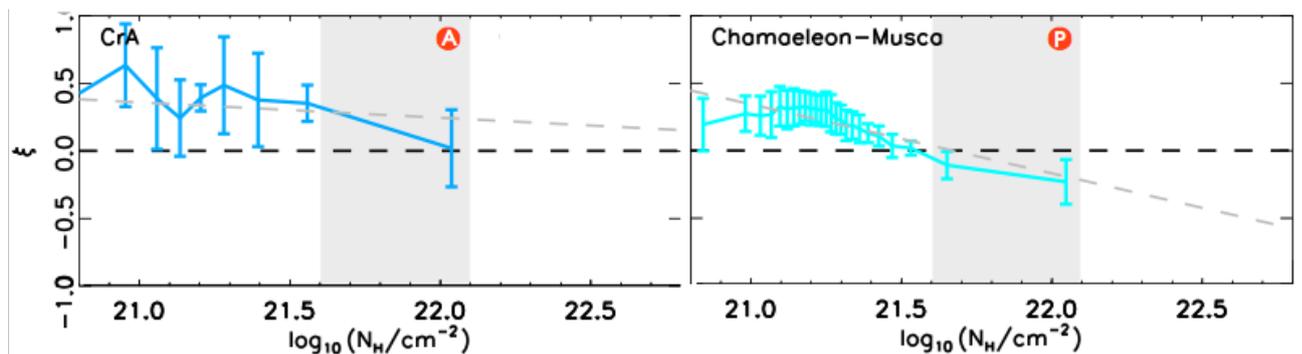

**Supplementary Figure 5 - Examples of individual clouds in Planck contour-field alignment analysis**[30]. ξ, adopted from reference 30, is the alignment parameter for column density contours and B-fields: a positive/negative ξ indicates an alignment closer to parallel/perpendicular. A trend of moving away from parallelism with increasing column density is usually seen. To compare with the cloud-field alignment in Figure 2, recall that Figure 2 focuses on $A_V > 2$ mag (i.e. where most of the YSOs distribute (Supplementary Figure 1)), which is indicated by the shaded regions. As the two plots show, the high-density regions can be either aligned with (Ⓐ) or perpendicular to (Ⓟ) the B-field. The two clouds are shown in Figure 2 with the same symbols.

**Supplementary Table 1** Mostly the same as Table 1 except that the B fields are based on Planck data.

| Cloud Name | Long axes[a] degrees | B fields[b] degrees | SFR/cloud mass % Myr$^{-1}$ | SFR/cloud mass % Myr$^{-1}$ |
|---|---|---|---|---|
| 1. IC 5146 | -38 | 67 ±14 | | 0.380±0.18 |
| 2. Pipe Nebula | -45 | 59 ±16 | 2.81 | |
| 3. Orion B | -30 | 93± 20 | 2.19 | |
| 4. Chamaeleon | 22 | -76± 13 | | 1.03±0.48 |
| 5. Taurus | 75 | 1 ± 27 | 4.76 | 0.140 |
| 6. Lupus I | -1 | 86 ±37 | 4.00 | 0.630±0.52 |
| 7. Musca | 27 | -82 ±8 | | 0.900±0.56 |
| 8. Perseus | 32 | -84 ±28 | 7.98 | 1.46±1.1 |
| 9. Lupus II-VI | -73 | 82 ± 18 | 6.97 | 1.85±0.74 |
| 10. Corona Aus. | -26 | -26 ± 24 | 9.69 | 3.66±2.4 |
| 11. Cepheus | 65 | -89 ± 14 | | 1.13±0.62 |
| 12 Ophiuchus | -45 | -81± 18 | 6.10 | 2.32±1.8 |
| 13. Aquila | -75 | -67 ± 17 | | 1.48±0.8 |
| 14. Orion A | 83 | 59 ± 25 | 5.21 | |

a. The SFR of Orion A and B are available individually but are combined in Table 1 due to reference 5 combining the two to fit the resolution of optical polarimetry data. The same applies to the pair of Chamaeleon and Musca. Given the higher resolution of Planck polarimetry data, these four regions are treated separately here. Their cloud orientations are obtained by applying the method proposed in reference 5 (see the main text) on the Planck $\tau_{353}$ data[30].

b. The mean directions are from the regions above the intermediate-density thresholds defined by Planck Collaboration[30]. See Supplementary Figure 1-3 for the maps of the selected areas. The uncertainties shown here are the interquartile ranges.